# Notes on approaches for solving the Euler-Poisson equations.

**Sergey V. Ershkov***, Affiliation: Plekhanov Russian University of Economics, Scopus number 60030998, e-mail: sergej-ershkov@yandex.ru,

**Dmytro Leshchenko**, Odessa State Academy of Civil Engineering and Architecture, Odessa, Ukraine, e-mail: leshchenko_d@ukr.net

## Abstract

In this paper, we proceed to develop a new approach which was formulated first in *Ershkov* (2017) for solving Poisson equations: a new type of the solving procedure for Euler-Poisson equations (rigid body rotation over the fixed point) is suggested in the current research. Meanwhile, the Euler-Poisson system of equations has been successfully explored for the existence of analytical way for presentation of the solution.

As the main result, the new ansatz is suggested for solving Euler-Poisson equations: the Euler-Poisson equations are reduced to the system of 3 nonlinear ordinary differential equations of 1-st order in regard to 3 functions $\Omega_i$ ($i$ = 1, 2, 3); the elegant *approximate* solution has been obtained via *re-inversion* of the proper *analytical* integral as a set of quasi-periodic cycles.

So, the system of Euler-Poisson equations is proved to have the analytical solutions (in quadratures) only in classical simplifying cases: 1) *Lagrange's* case, or 2) *Kovalevskaya's* case or 3) *Euler's* case or other well-known but particular cases.

## 1. Introduction, equations of motion.

Euler-Poisson equations, describing the dynamics of rigid body rotation, are known to be one of the famous problems in classical mechanics.

In accordance with [1-3], Euler equations describe the rotation of a rigid body in a frame of reference fixed in the rotating body for the case of rotation over the fixed point as below (*at given initial conditions*):

$$\begin{cases} I_1 \dfrac{d\,\Omega_1}{d\,t} + (I_3 - I_2)\,\Omega_2\,\Omega_3 = P\left(\gamma_2 c_0 - \gamma_3 b_0\right), \\[2ex] I_2 \dfrac{d\,\Omega_2}{d\,t} + (I_1 - I_3)\,\Omega_3\,\Omega_1 = P\left(\gamma_3 a_0 - \gamma_1 c_0\right), \\[2ex] I_3 \dfrac{d\,\Omega_3}{d\,t} + (I_2 - I_1)\,\Omega_1\,\Omega_2 = P\left(\gamma_1 b_0 - \gamma_2 a_0\right), \end{cases} \tag{1.1}$$

- where $I_i \neq 0$ are the principal moments of inertia ($i$ = 1, 2, 3) and $\Omega_i$ are the components of the *angular velocity vector* along the proper principal axis; $\gamma_i$ are the components of the weight of mass $P$ and $a_0$, $b_0$, $c_0$ are the appropriate coordinates of the center of masses in a frame of reference fixed in the rotating body (*in regard to the absolute system of coordinates X, Y, Z*).

Poisson equations for the components of the weight in a frame of reference fixed in the rotating body (*in regard to the absolute system of coordinates X, Y, Z*) should be presented as below [4-5]:

$$\begin{cases} \dfrac{d\gamma_1}{dt} = \Omega_3\gamma_2 - \Omega_2\gamma_3\,, \\[2ex] \dfrac{d\gamma_2}{dt} = \Omega_1\gamma_3 - \Omega_3\gamma_1\,, \\[2ex] \dfrac{d\gamma_3}{dt} = \Omega_2\gamma_1 - \Omega_1\gamma_2\,, \end{cases} \qquad (1.2)$$

besides, we should present the invariants (*first integrals of motion*) as below

$$\begin{cases} \gamma_1^2 + \gamma_2^2 + \gamma_3^2 = 1\,, \\[2ex] I_1\Omega_1\gamma_1 + I_2\Omega_2\gamma_2 + I_3\Omega_3\gamma_3 = const = C_0\,, \\[2ex] \dfrac{1}{2}\left(I_1\Omega_1^2 + I_2\Omega_2^2 + I_3\Omega_3^2\right) + P\,(a_0\gamma_1 + b_0\gamma_2 + c_0\gamma_3) = const = C_1\,. \end{cases} \qquad (1.3)$$

Whereas the 3-rd and 2-nd invariant of (1.3) are proved to be fully reflecting all the dynamical properties of the Euler-Poisson system (1.1)-(1.2) [6], the 1-st integral of (1.3) belongs to Poisson equations only: to obtain it, we should multiply each of equations of (1.2) on $\gamma_i$ accordingly, then sum it to each other (*the constant of integration is chosen equal to 1, due to trigonometric sense of the presenting solution in absolute system of coordinates via Euler angles*)

$$\frac{1}{2}\frac{d}{dt}(\gamma_1{}^2) + \frac{1}{2}\frac{d}{dt}(\gamma_2{}^2) + \frac{1}{2}\frac{d}{dt}(\gamma_3{}^2) = 0\,,$$

As we can see, only 2 of 3 proper additional invariants (1.3) above are obtained by using of all the 6 EP-equations (including Poisson equations).

This is not sufficient for the trying to solve EP-equations: indeed, system of equations (1.1)-(1.2) is supposed *not to be equivalent* to the system of equations (1.1) along with all the invariants (1.3) (*Prof. D. Sc. Hamad Yehia, Mansoura University, Egypt, personal communications*) for some particular cases, as it was suggested earlier in [7-8] (the rather complex case was considered in [9]). So, for solving system of equations (1.1)-(1.2), we should first solve the Poisson equations (1.2).

## 2. Presentation of the solution of Poisson equations.

The system of Eqs. (1.2) has *the analytical* way to present the general solution [6] (in regard to the time-parameter *t*), in fluid mechanics see refs. [10]-[13]

$$\gamma_1 = -\sigma\left(\frac{2a}{1+(a^2+b^2)}\right), \quad \gamma_2 = -\sigma\left(\frac{2b}{1+(a^2+b^2)}\right),$$

$$\gamma_3 = \sigma\left(\frac{1-(a^2+b^2)}{1+(a^2+b^2)}\right), \qquad (2.1)$$

where $\sigma$ is some arbitrary (real) constant, given by the initial conditions; $\Omega_i$ are the functions of time-parameter *t* only - it means that $\Omega_i \neq \Omega_i(\{\gamma_i\}, t)$.

Besides, the real-valued coefficients $a(t)$, $b(t)$ (2.1) are solutions of the mutual system of 2 *Riccati* ordinary differential equations:

$$\begin{cases} a' = \dfrac{\Omega_2}{2}a^2 - (\Omega_1 b)a - \dfrac{\Omega_2}{2}(b^2-1) + \Omega_3 b\,, \\[2ex] b' = -\dfrac{\Omega_1}{2}b^2 + (\Omega_2 a)b + \dfrac{\Omega_1}{2}(a^2-1) - \Omega_3 a\,. \end{cases} \qquad (2.2)$$

Equations (2.2) above are proved to be the classical *Riccati* ODEs [6]. Each of them describes the evolution of function $a(t)$ in dependence on the function $b(t)$ (and *vice versa*) along with the functions $\{\Omega_i\}$ in regard to the time *t*; aforementioned *Riccati* ODE has no analytical solution in general case [14].

## 3. Presentation of the solution of Euler-Poisson equations.

According to the results of previous section, the Poisson system of Eqs. (1.2) has *the analytical* way to present its general solution in the form (2.1). Meanwhile, as we can see from Section 1 above, two of three proper additional invariants (1.3) are obtained by using of all the 6 Euler-Poisson equations (1.1)-(1.2).

Thus, we can make a reasonable conclusion that system of equations (1.1)-(1.2) is supposed *to be equivalent* to the system of *Poisson* equations (1.2) along with updated *Euler* equations (1.1): to any two of them the last two invariants of (1.3) could be substituted by (for the theory of invariants of ODE-systems, see [14]).

So, for solving *Euler-Poisson* system of equations (1.1)-(1.2), we should first solve the *Poisson* equations (1.2) in a form (2.1)-(2.2), which should be accomplished with the two aforementioned invariants along with any 1 equation of 3 equations (1.1) (for definiteness, let us choose the 3-rd equation from Eqns. (1.1)):

$$
\begin{cases}
\begin{cases}
I_1\Omega_1\gamma_1 + I_2\Omega_2\gamma_2 + I_3\Omega_3\gamma_3 = const = C_0\,, \\
\frac{1}{2}\left(I_1\Omega_1^2 + I_2\Omega_2^2 + I_3\Omega_3^2\right) + P(a_0\gamma_1 + b_0\gamma_2 + c_0\gamma_3) = const = C_1\,, \\
I_3\frac{d\Omega_3}{dt} + (I_2 - I_1)\Omega_1\Omega_2 = P\left(\gamma_1 b_0 - \gamma_2 a_0\right),
\end{cases} \\
\\
\gamma_1 = -\sigma\left(\frac{2a}{1+(a^2+b^2)}\right), \quad \gamma_2 = -\sigma\left(\frac{2b}{1+(a^2+b^2)}\right), \quad \gamma_3 = \sigma\left(\frac{1-(a^2+b^2)}{1+(a^2+b^2)}\right), \qquad (3.1) \\
\Updownarrow \\
\left\{\Leftrightarrow \gamma_1^2 + \gamma_2^2 + \gamma_3^2 = 1\right\} \\
\begin{cases}
a' = \frac{\Omega_2}{2}a^2 - (\Omega_1 b)a - \frac{\Omega_2}{2}(b^2-1) + \Omega_3 b\,, \\
b' = -\frac{\Omega_1}{2}b^2 + (\Omega_2 a)b + \frac{\Omega_1}{2}(a^2-1) - \Omega_3 a\,.
\end{cases}
\end{cases}
$$

Using the expressions for $\gamma_i$ in (3.1), we obtain from 1-st and 2-nd equation of system (3.1) as below:

$$\begin{cases} I_1\Omega_1\left(-\sigma\left(\frac{2a}{1+(a^2+b^2)}\right)\right)+I_2\Omega_2\left(-\sigma\left(\frac{2b}{1+(a^2+b^2)}\right)\right)+I_3\Omega_3\,\sigma\left(\frac{2}{1+(a^2+b^2)}-1\right)=C_0, \\ \frac{1}{2P}\left(I_1\Omega_1^2+I_2\Omega_2^2+I_3\Omega_3^2\right)+a_0\left(-\sigma\left(\frac{2a}{1+(a^2+b^2)}\right)\right)+b_0\left(-\sigma\left(\frac{2b}{1+(a^2+b^2)}\right)\right)+c_0\,\sigma\left(\frac{2}{1+(a^2+b^2)}-1\right)=\frac{C_1}{P}, \end{cases}$$

or

$$\begin{cases} -2I_1\Omega_1\frac{a\sigma}{1+(a^2+b^2)}-2I_2\Omega_2\frac{b\sigma}{1+(a^2+b^2)}+2I_3\Omega_3\frac{\sigma}{1+(a^2+b^2)}=C_0+I_3\Omega_3\sigma, \\ -2a_0\frac{a\sigma}{1+(a^2+b^2)}-2b_0\frac{b\sigma}{1+(a^2+b^2)}+2c_0\frac{\sigma}{1+(a^2+b^2)}=\frac{C_1}{P}+c_0\sigma-\frac{1}{2P}\left(I_1\Omega_1^2+I_2\Omega_2^2+I_3\Omega_3^2\right). \end{cases} \qquad (3.2)$$

Let us divide the left part of 2-nd equation (3.2) above onto the left part of the first equation (accomplishing also such the dividing with respect to the right parts for both of these equations) as below

$$\frac{a_0\frac{a\sigma}{1+(a^2+b^2)}+b_0\frac{b\sigma}{1+(a^2+b^2)}-c_0\frac{\sigma}{1+(a^2+b^2)}}{I_1\Omega_1\frac{a\sigma}{1+(a^2+b^2)}+I_2\Omega_2\frac{b\sigma}{1+(a^2+b^2)}-I_3\Omega_3\frac{\sigma}{1+(a^2+b^2)}}=\frac{\frac{C_1}{P}+c_0\sigma-\frac{1}{2P}\left(I_1\Omega_1^2+I_2\Omega_2^2+I_3\Omega_3^2\right)}{C_0+I_3\Omega_3\sigma},$$

$\Rightarrow$

$$\frac{a_0a+b_0b-c_0}{I_1\Omega_1a+I_2\Omega_2b-I_3\Omega_3}=\frac{\frac{C_1}{P}+c_0\sigma-\frac{1}{2P}\left(I_1\Omega_1^2+I_2\Omega_2^2+I_3\Omega_3^2\right)}{C_0+I_3\Omega_3\sigma}, \quad \Rightarrow$$

$$a = b\left(\frac{\Upsilon I_2\Omega_2 - b_0}{a_0 - \Upsilon I_1\Omega_1}\right) + \left(\frac{c_0 - \Upsilon I_3\Omega_3}{a_0 - \Upsilon I_1\Omega_1}\right), \quad \left\{ \Upsilon = \frac{\frac{C_1}{P} + c_0\sigma - \frac{1}{2P}\left(I_1\Omega_1^2 + I_2\Omega_2^2 + I_3\Omega_3^2\right)}{C_0 + I_3\Omega_3\sigma} \right\} \qquad (3.3)$$

So, the last equation (3.3) reveals the linear dependence of function $a(t)$ with respect to the function $b(t)$ for solutions of system (3.1). In addition to the aforesaid linear invariant (3.3), we obtain from the first of equations (3.2)

$$F b^2 + 2Gb + H = 0, \quad \Rightarrow \quad b = \frac{-G \pm \sqrt{G^2 - FH}}{F}, \qquad (3.4)$$

$$F = \left(\frac{C_0}{\sigma} + I_3\Omega_3\right)\left(\left(\frac{\Upsilon I_2\Omega_2 - b_0}{a_0 - \Upsilon I_1\Omega_1}\right)^2 + 1\right),$$

$$G = \left(\left(\frac{C_0}{\sigma} + I_3\Omega_3\right)\left(\frac{\Upsilon I_2\Omega_2 - b_0}{a_0 - \Upsilon I_1\Omega_1}\right)\cdot\left(\frac{c_0 - \Upsilon I_3\Omega_3}{a_0 - \Upsilon I_1\cdot\Omega_1}\right) + I_1\Omega_1\left(\frac{\Upsilon I_2\Omega_2 - b_0}{a_0 - \Upsilon I_1\Omega_1}\right) + I_2\Omega_2\right),$$

$$H = \left(\frac{C_0}{\sigma} + I_3\Omega_3\right)\left(\left(\frac{c_0 - \Upsilon I_3\Omega_3}{a_0 - \Upsilon I_1\Omega_1}\right)^2 + 1\right) + 2I_1\Omega_1\left(\frac{c_0 - \Upsilon I_3\Omega_3}{a_0 - \Upsilon I_1\Omega_1}\right) - 2I_3\Omega_3$$

The last but not least, we should especially note that system of equations (3.1) is reduced to the system (3.4)-(3.5) of 3 nonlinear ordinary differential equations of the 1-st order in regard to 3 functions $\Omega_i$ ($i$ = 1, 2, 3):

$$\begin{cases} I_3\dfrac{d\Omega_3}{dt} + (I_2 - I_1)\Omega_1\Omega_2 = P\left(\gamma_1 b_0 - \gamma_2 a_0\right), \\ a' = \dfrac{\Omega_2}{2}a^2 - (\Omega_1 b)a - \dfrac{\Omega_2}{2}(b^2 - 1) + \Omega_3 b, \\ b' = -\dfrac{\Omega_1}{2}b^2 + (\Omega_2 a)b + \dfrac{\Omega_1}{2}(a^2 - 1) - \Omega_3 a, \end{cases} \qquad (3.5)$$

where

$$\gamma_1 = -\sigma\left(\frac{2a}{1+(a^2+b^2)}\right),\quad \gamma_2 = -\sigma\left(\frac{2b}{1+(a^2+b^2)}\right),\quad \gamma_3 = \sigma\left(\frac{1-(a^2+b^2)}{1+(a^2+b^2)}\right),$$

$$a = b\left(\frac{\Upsilon I_2\,\Omega_2 - b_0}{a_0 - \Upsilon I_1\Omega_1}\right) + \left(\frac{c_0 - \Upsilon I_3\Omega_3}{a_0 - \Upsilon I_1\Omega_1}\right),\quad \left\{\Upsilon = \frac{\frac{C_1}{P} + c_0\sigma - \frac{1}{2P}\left(I_1\Omega_1^2 + I_2\Omega_2^2 + I_3\Omega_3^2\right)}{C_0 + I_3\Omega_3\sigma}\right\},$$

$$b = \frac{-G \pm \sqrt{G^2 - FH}}{F},$$

$$\left\{\begin{array}{l} F = (\frac{C_0}{\sigma} + I_3\Omega_3)\left(\left(\frac{\Upsilon I_2\,\Omega_2 - b_0}{a_0 - \Upsilon I_1\Omega_1}\right)^2 + 1\right), \\ G = \left((\frac{C_0}{\sigma} + I_3\Omega_3)\left(\frac{\Upsilon I_2\Omega_2 - b_0}{a_0 - \Upsilon I_1\Omega_1}\right)\left(\frac{c_0 - \Upsilon I_3\Omega_3}{a_0 - \Upsilon I_1\Omega_1}\right) + I_1\Omega_1\left(\frac{\Upsilon I_2\Omega_2 - b_0}{a_0 - \Upsilon I_1\Omega_1}\right) + I_2\Omega_2\right), \\ H = (\frac{C_0}{\sigma} + I_3\Omega_3)\left(\left(\frac{c_0 - \Upsilon I_3\Omega_3}{a_0 - \Upsilon I_1\Omega_1}\right)^2 + 1\right) + 2I_1\Omega_1\left(\frac{c_0 - \Upsilon I_3\Omega_3}{a_0 - \Upsilon I_1\Omega_1}\right) - 2I_3\Omega_3 \end{array}\right\}$$

## 4. <u>The case of approximate solution of Euler-Poisson equations.</u>

Let us choose the simplifying assumption for solving system of equations (3.5) (for example, assumption of *symmetric* rigid rotor):

$$I_1 = I_2,\quad \{a_0, b_0\} = 0 \qquad (4.1)$$

So, let us consider the case of symmetric rigid rotor (4.1) where its center of mass lies on the symmetry axis. It is obviously similar to the Lagrange case, but in Lagrange case we additionally assume [1]:

1) the angular momentum component along the symmetry axis,
2) the angular momentum in the Z-direction and
3) the magnitude of the $\gamma$-vector

whereas all of the aforesaid invariants should be constant.

If we assume (4.1) such simplifications should transform the 1-st equation of (3.5) to the simple equality $\Omega_3$ = const. As we can see from the structure of components of equations (3.5), we still need additional simplifying assumption to solve it easily:

$$|I_1 \Omega_1| \gg \{|I_2 \Omega_2|, |I_3 \Omega_3|\} \qquad (4.2)$$

Conditions (4.2) can be associated with evolution of spin of the rotating rigid body from initial-current spin state towards the rotation about maximal-inertia axis due to the process of *nutation relaxation* or to the proper spin state corresponding to minimal energy (which is fluctuating near the given appropriate constant) with approximately fixed angular momentum.

Let us also consider an additional simplifying assumption as below ($\Omega_3$ = const):

$$c_0 \rightarrow \Upsilon I_3 \Omega_3 , \qquad (4.3)$$

where

$$\Upsilon = \frac{\dfrac{C_1}{P} + c_0 \sigma - \dfrac{1}{2P}\left(I_1 \Omega_1^2 + I_2 \Omega_2^2 + I_3 \Omega_3^2\right)}{C_0 + I_3 \Omega_3 \sigma}$$

whereas the expression below

$$\left(I_1 \Omega_1^2 + I_2 \Omega_2^2 + I_3 \Omega_3^2\right)$$

is associated with the approximately fixed angular momentum (here we refer to the aforementioned evolution of rigid body rotation towards the spinning over maximal-inertia axis due to the process of *nutation relaxation*).

Meanwhile, the aforesaid assumptions (4.1)-(4.3) should simplify components of the system of equations (3.5) accordingly:

$$a = -b\left(\frac{\Omega_2}{\Omega_1}\right) - \left(\frac{c_0 - \Upsilon I_3\Omega_3}{\Upsilon I_1\Omega_1}\right), \quad b = \frac{-G \pm \sqrt{G^2 - FH}}{F}, \quad I_3\Omega_3 \geq \left(\frac{C_0}{\sigma}\right)$$

$$\left\{\begin{aligned} F &= (\frac{C_0}{\sigma} + I_3\Omega_3)\left(\left(\frac{\Omega_2}{\Omega_1}\right)^2 + 1\right) \cong (\frac{C_0}{\sigma} + I_3\Omega_3), \\ G &= (\frac{C_0}{\sigma} + I_3\Omega_3)\left(\frac{\Omega_2}{\Omega_1}\right)\left(\frac{c_0 - \Upsilon I_3\Omega_3}{\Upsilon I_1\Omega_1}\right) \cong 0, \\ H &= (\frac{C_0}{\sigma} + I_3\Omega_3)\left(\left(\frac{c_0 - \Upsilon I_3\Omega_3}{\Upsilon I_1\Omega_1}\right)^2 + 1\right) - \left(\frac{2c_0}{\Upsilon}\right) \cong (\frac{C_0}{\sigma} - I_3\Omega_3) \end{aligned}\right\}$$

which therefore should be presented as below

$$a = -b\left(\frac{\Omega_2}{\Omega_1}\right), \quad b = \pm\sqrt{\frac{(I_3\Omega_3)^2 - (\frac{C_0}{\sigma})^2}{\left(I_3\Omega_3 + \frac{C_0}{\sigma}\right)^2}} \qquad (4.4)$$

Conditions (4.4) above imply that function $a(t)$ depends on both the components $\Omega_1$, $\Omega_2$ (as first approximation), whereas function $b(t)$ is to be appropriate constant.

It helps us in simplifying the third equation of system (3.5) (mathematical procedure of simplifying has been moved to an **Appendix A1**, with only resulting formulae left in the main text):

$$\Omega_1 = \sqrt{\frac{2b(\Omega_2 \Omega_3) - (\Omega_2)^2 b^2}{(b^2 + 1)}} \ , \qquad (4.5)$$

$$\Omega_2 \le \frac{2b\Omega_3}{b^2}$$

and also it helps us in simplifying the second equation of system (3.5) accordingly (see also **Appendix A1**):

$$-\frac{d}{dt}\left(\frac{\Omega_2}{\Omega_1}\right) = \frac{\Omega_2}{2} b \left(\frac{\Omega_2}{\Omega_1}\right)^2 + \frac{\Omega_2}{2} b + \frac{\Omega_2}{2b} + \Omega_3 \qquad (4.6)$$

where expression for $\Omega_2$ should be determined from the Eqn. (4.5) above, but Eqn. (4.6) yields as below (all the simple algebra manipulations are moved also to the **Appendix A1**):

$$\left(\frac{d\Omega_2}{(\Omega_2 + 2b\Omega_3)\sqrt{2(\Omega_2 \Omega_3) - (\Omega_2)^2 b}}\right) = -\frac{1}{b}\sqrt{\frac{b}{(b^2+1)}}\, dt \qquad (4.7)$$

The left side of equation (4.7) could be transformed to the proper *analytical* integral [10] in regard to function $\Omega_2$ [15-16] (let us choose the sign "+" in formulae (4.4) for *b*; for simple algebra manipulations, see **Appendix A1**):

$$\int \frac{d\Omega_2}{(\Omega_2 + 2b\Omega_3)\sqrt{2(\Omega_2 \Omega_3) - (\Omega_2)^2 b}} = \frac{\arcsin\left(\frac{(1+2b^2)\Omega_2 - 2b\Omega_3}{(\Omega_2 + 2b\Omega_3)}\right)}{2\Omega_3\sqrt{b+b^3}} \qquad (4.8)$$

Thus, by the obtaining of *re-inverse* dependence for the expression (4.8), we could present the solution as a set of quasi-periodic cycles: - it means a quasi-periodic character of the evolution of components $\Omega_2$ and $\Omega_1$ (4.5) of angular velocity of rigid body rotation

$$\Omega_2 = 2\Omega_3 \left( \frac{b - b\sin(2\Omega_3 t)}{1 + 2b^2 + \sin(2\Omega_3 t)} \right) \tag{4.9}$$

## 5. <u>Presentation of approximate solution (4.5)-(4.9), via *Euler's* angles.</u>

According to the results in [7], Euler's kinematic equations, which describe the rotation of a rigid body over the fixed point in regard to the fixed Cartesian coordinate system, should be presented as below (*at given initial conditions*):

$$\begin{cases} \Omega_1 = \dot{\psi}\gamma_1 + \dot{\theta}\cos\varphi, \\ \Omega_2 = \dot{\psi}\gamma_2 - \dot{\theta}\sin\varphi, \\ \Omega_3 = \dot{\psi}\gamma_3 + \dot{\varphi}, \end{cases} \tag{5.1}$$

$$\begin{cases} \gamma_1 = \sin\theta\sin\varphi, \\ \gamma_2 = \sin\theta\cos\varphi, \\ \gamma_3 = \cos\theta \end{cases} \tag{5.2}$$

where $\psi$, $\theta$, $\varphi$ - are the appropriate angles, describing the positions of the reference fixed in the rotating body (*in regard to the absolute system of coordinates X*, *Y*, *Z*), see Fig.1:

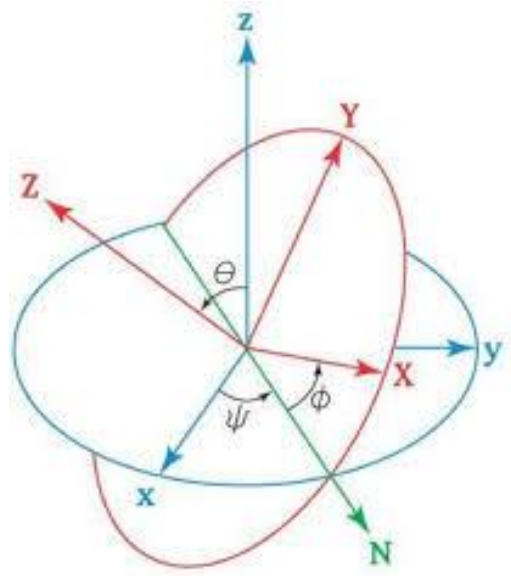

Fig. 1. Presentation of Euler's angles.

Equations (5.1)-(5.2) along with expressions for approximate solution (4.4)-(4.5), (4.9) let us obtain as below (taking into account also expressions (3.1) for $\gamma_i$):

$$\begin{cases} \varphi = \arctan(\gamma_1/\gamma_2), \\ \theta = \arccos\gamma_3 \end{cases} \Rightarrow \begin{cases} \varphi = \arctan\left(\dfrac{a}{b}\right), \\ \theta = \arccos\left(\sigma\left(\left(\dfrac{2}{1+(a^2+b^2)}\right)-1\right)\right), \end{cases} \tag{5.3}$$

which could be presented as below

$$\frac{a}{b} = -\frac{\Omega_2}{\sqrt{\dfrac{2b\Omega_2\Omega_3-(\Omega_2)^2 b^2}{(b^2+1)}}} = -\sqrt{\frac{(b^2+1)}{2b\dfrac{\Omega_3}{\Omega_2}-b^2}} = -\sqrt{\frac{1-\sin(2\Omega_3 t)}{1+\sin(2\Omega_3 t)}},$$

$$\left(\left(\frac{2}{1+(a^2+b^2)}\right)-1\right) = \frac{2}{\left(1+b^2\left(\dfrac{1-\sin(2\Omega_3 t)}{1+\sin(2\Omega_3 t)}\right)+b^2\right)} - 1 = \left(\frac{1+\sin(2\Omega_3 t)-2b^2}{1+\sin(2\Omega_3 t)+2b^2}\right) \Rightarrow$$

$$\Rightarrow \begin{cases} \varphi = -\arctan\left(\sqrt{\dfrac{1-\sin(2\Omega_3 t)}{1+\sin(2\Omega_3 t)}}\right), \\ \theta = \arccos\left(\sigma\left(\dfrac{1+\sin(2\Omega_3 t)-2b^2}{1+\sin(2\Omega_3 t)+2b^2}\right)\right), \end{cases} \tag{5.4}$$

Furthermore, using (5.4), the appropriate expression for the dynamics of angle $\psi$ could be obtained from one of the Eqns. (5.1) (for simple algebra manipulations, see **Appendix A2**):

$$\dot{\psi} = \frac{\Omega_3 - \dot{\varphi}}{\gamma_3} \quad \Rightarrow \quad \dot{\psi} = \frac{\Omega_3 + \dfrac{d}{dt}\left(\arctan\left(\sqrt{\dfrac{1-\sin(2\Omega_3 t)}{1+\sin(2\Omega_3 t)}}\right)\right)}{\sigma\left(\dfrac{1+\sin(2\Omega_3 t) - 2b^2}{1+\sin(2\Omega_3 t) + 2b^2}\right)} \quad \Rightarrow$$

$$\Rightarrow \qquad \dot{\psi} = 0 \quad \Rightarrow \quad \psi = const \qquad (5.5)$$

Thus, formulae (5.4)-(5.5) are proved to describe the appropriate dynamics of rigid body rotation for the case of approximate solution (4.4)-(4.5), (4.9) in regard to the absolute system of coordinates *X*, *Y*, *Z*, via *Euler's* angles.

## 6. <u>Discussion.</u>

System of Euler-Poisson equations (which governs by the dynamics of rigid body rotation over fixed point), is known to be very hard to solve analytically: the aforementioned system is proved to have the analytical solutions (in quadratures) only in classical simplifying cases [18]: 1) *Lagrange's* case, or 2) *Kovalevskaya's* case or 3) *Euler's* case or other well-known but particular cases (where the existence of particular solutions depends on the choosing of the appropriate initial conditions).

A new approach is developed previously in [6] for solving of the Poisson equations in case the components of angular velocity of rigid body rotation could be considered as the functions of time-parameter *t* only. Fundamental solution is presented by the analytical formulae in dependence on two time-dependent, the real-valued coefficients. Such coefficients as above are proved to be the solutions of mutual system of 2 *Riccati* ordinary differential equations (which has no analytical solution in general case).

We proceed the exploring of the ansatz, which was formulated in [6] at first for solving of Poisson equations: the Euler-Poisson system of equations was explored here for the existence of analytical way for presentation of the solution.

As we can see from Section 1 above ("Introduction, equations of motion"), two of three the proper additional invariants (1.3) of EP-system of equations are obtained by using of all the 6 Euler-Poisson equations (1.1)-(1.2).

Thus, we can make a reasonable conclusion that system of equations (1.1)-(1.2) is supposed *to be equivalent* to the system of *Poisson* equations (1.2) along with updated *Euler* equations (1.1): to any two of them the last two invariants of (1.3) could be substituted by.

So, for solving Euler-Poisson system of equations (1.1)-(1.2), we should first solve the Poisson equations (1.2) in a form (2.1)-(2.2), which should be accomplished with the two aforementioned invariants along with any 1 equation of 3 equations (1.1) (for example, let us choose the 3-rd equation from Eqns. (1.1) for definiteness).

Having solved them according to the aforementioned ansatz, we should especially note that system of equations (3.1) is reduced to the system (3.5) of 3 nonlinear ordinary differential equations of 1-st order in regard to 3 functions $\Omega_i$ ($i$ = 1, 2, 3).

In our derivation, the main motivation was the transforming of the presented system of equations (3.5) to the convenient form, in which the minimum of numerical calculations are required to obtain the final solutions. Preferably, it should be the analytical or semi-analytical solutions; we have presented here the elegant *approximate* solution as a set of quasi-periodic cycles via re-inversion of the proper *analytical* integral.

Let us schematically imagine at Figs.2-3 the appropriate dynamics of angles θ, φ (depending on time $t$), according to the formulae (5.4), which corresponds to the aforementioned *approximate* solution (4.4)-(4.5), (4.9). Here the aforementioned angle θ describes deviation of the rotating symmetric rigid body from vertical axis OZ (of *absolute system of coordinates X*, *Y*, *Z*), whereas the angle φ describes dynamics of rotation over the axis of symmetry of rigid body itself.

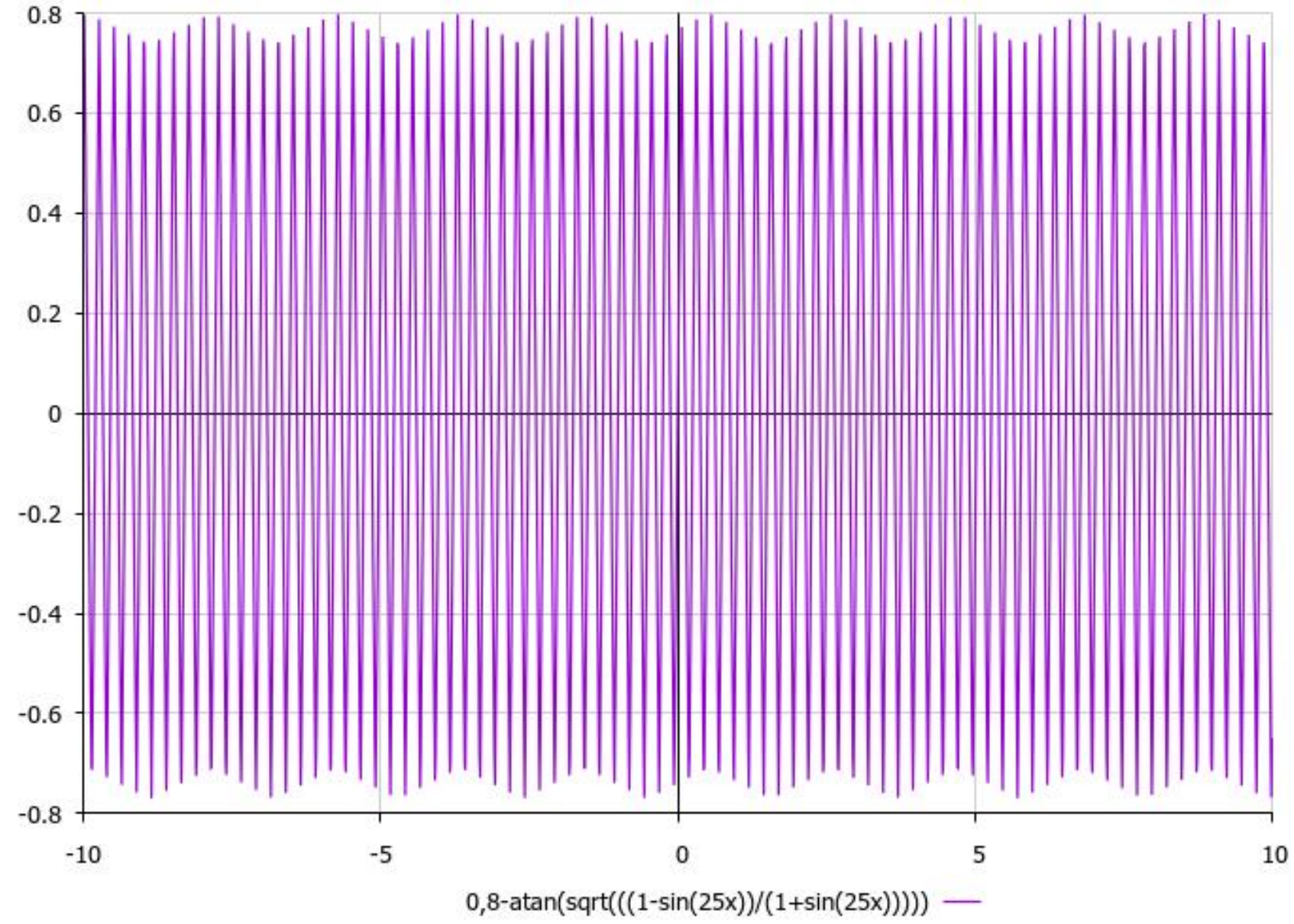

Fig.2. A *schematic* plot of angle θ, according to the formulae (5.4) (here we designate $x = t$ as argument on horizontal axis, just for the aim of presenting the plot of solution)

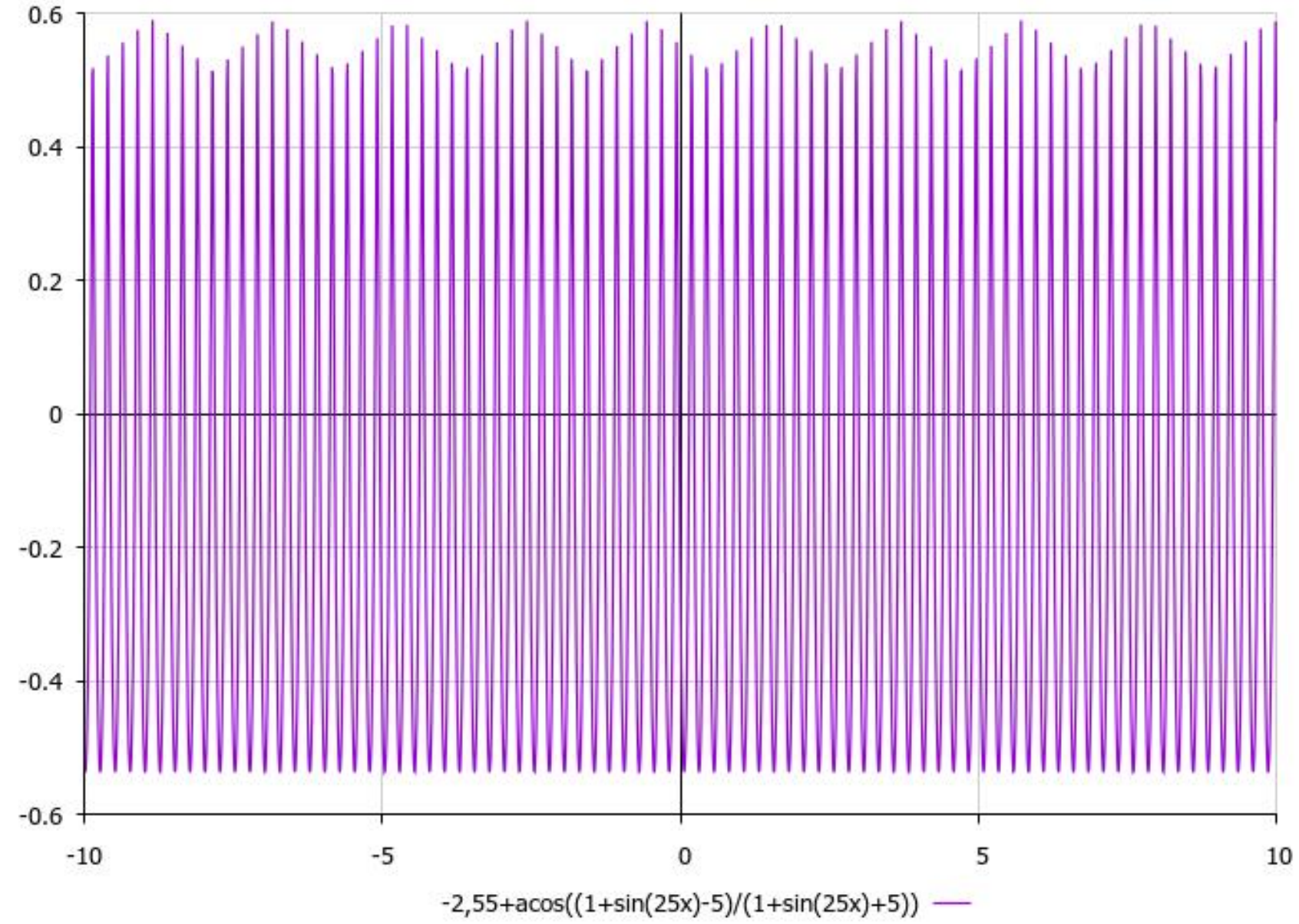

Fig.3. A *schematic* plot of angle φ, according to the formulae (5.4) (σ = 1).

## 7. **Conclusion, the short survey on recent issues in literature on rigid body rotation.**

A new approach was firstly formulated in [17] for solving of Poisson equations: the Euler-Poisson system of equations was explored there for the existence of analytical way for presentation of the solution. Main motivation of the aforementioned research was to correct the previous issue: indeed, the system of equations (1.1)-(1.2) is supposed *not to be equivalent* to the system of equations (1.1) along with all the invariants (1.3) for some particular cases, as it was suggested earlier in [7-8]. If you solve the dynamical equations (1.1) using only integrals (1.3) without Poisson equations (1.2), some untrue solutions of Euler-Poisson equations may come through [7-8].

As a result of thinking over the aforesaid issues in previous researches, a new ansatz was suggested for solving of Euler-Poisson equations in [6], which has been successfully developed in the current research: the Euler-Poisson equations are reduced to the system (3.5) of 3 nonlinear ODE (ordinary differential equations of 1-st order) in regard to 3 functions $\Omega_i$ ($i$ = 1,2,3); the proper elegant *approximate* solution has been obtained as a set of quasi-periodic cycles via re-inversion of the proper *analytical* integral.

The last but not least, we should note that despite the article [17] having been published first (moreover, this manuscript with self-criticism was received by Editorial Office at 25 September 2016), untrue solutions of Euler-Poisson equations [7-8] were also under criticism of work [19], which was published later than [17] (it was received by the Editorial Office of that journal at 12 March 2017).

It is more astonishing that two years later authors of [20] reported their results without correct citing of article [17], this fact is obviously against scientific ethic and can be considered as plagiarism.

Also, some remarkable articles should be cited, which concern the problem under consideration, [21]-[27].

## 8. Acknowledgements.

Sergey Ershkov is thankful to Prof. D. Sc. Hamad Yehia, Mansoura University, Egypt as for the insightful argumentation during mutual fruitful discussions in the process of preparing of the previous manuscript (which is the basis for the current research).

Remark regarding contributions of authors as below:

In this research, Sergey Ershkov is responsible for the general ansatz and the solving procedure, simple algebra manipulations, calculations, results of the article in Sections 1-6 and also is responsible for the search of approximate solutions (as well as for self-critical remarks in Section 7).
Dr. Dmytro Leshchenko is responsible for theoretical investigations as well as for the deep survey in literature on the problem under consideration.
Both authors agreed with the results and conclusions of each other in Sections 1-7.

## Appendix, A1 (simplifying the third equation of system (3.5) to the form (4.5)).

Let us simplify the third equation of system (3.5) by using formulae (4.4)

$$0=-\frac{\Omega_1}{2}b^2+(\Omega_2 a)b+\frac{\Omega_1}{2}(a^2-1)-\Omega_3 a\,,\quad b=\pm\sqrt{\frac{(I_3\Omega_3)^2-(\frac{C_0}{\sigma})^2}{\left(I_3\Omega_3+\frac{C_0}{\sigma}\right)^2}}$$

$$0=-\frac{1}{2}b^2-(\frac{\Omega_2}{\Omega_1}b\left(\frac{\Omega_2}{\Omega_1}\right))b+\frac{1}{2}(b^2\left(\frac{\Omega_2}{\Omega_1}\right)^2-1)+\frac{\Omega_3}{\Omega_1}b\left(\frac{\Omega_2}{\Omega_1}\right),\quad\Rightarrow$$

$$\Omega_1=\sqrt{\frac{2b(\Omega_2\Omega_3)-(\Omega_2)^2 b^2}{(b^2+1)}}\;, \qquad (4.5)$$

$$\Omega_2\le\frac{2b\Omega_3}{b^2}$$

and also the aforementioned formulae (4.4) help us in simplifying the second equation of system (3.5) accordingly:

$$\frac{d}{dt}\left(-b\left(\frac{\Omega_2}{\Omega_1}\right)\right)=\frac{\Omega_2}{2}\left(b\left(\frac{\Omega_2}{\Omega_1}\right)\right)^2+(\Omega_1 b)b\left(\frac{\Omega_2}{\Omega_1}\right)-\frac{\Omega_2}{2}(b^2-1)+\Omega_3 b\quad\Rightarrow$$

$$-\frac{d}{dt}\left(\frac{\Omega_2}{\Omega_1}\right)=\frac{\Omega_2}{2}b\left(\frac{\Omega_2}{\Omega_1}\right)^2+\frac{\Omega_2}{2}b+\frac{\Omega_2}{2b}+\Omega_3 \qquad (4.6)$$

where expression for $\Omega_2$ should be determined from the Eqn. (4.5) above, but Eqn. (4.6) yields as below

$$\left( \frac{\sqrt{\frac{2b(\Omega_2\Omega_3)-(\Omega_2)^2 b^2}{(b^2+1)}} - \frac{\left(\frac{b\Omega_3-\Omega_2 b^2}{(b^2+1)}\right)}{\sqrt{\frac{2b(\Omega_2\Omega_3)-(\Omega_2)^2 b^2}{(b^2+1)}}}(\Omega_2)}{\left(\frac{2b(\Omega_2\Omega_3)-(\Omega_2)^2 b^2}{(b^2+1)}\right)} \right) \frac{d}{dt}(\Omega_2) =$$

$$= -\frac{\Omega_2}{2} b \frac{(\Omega_2)^2}{\left(\frac{2b(\Omega_2\Omega_3)-(\Omega_2)^2 b^2}{(b^2+1)}\right)} - \frac{\Omega_2}{2} b - \frac{\Omega_2}{2b} - \Omega_3 \qquad \Rightarrow$$

$$\left( \frac{d\Omega_2}{(\Omega_2+2b\Omega_3)\sqrt{2(\Omega_2\Omega_3)-(\Omega_2)^2 b}} \right) = -\frac{1}{b}\sqrt{\frac{b}{(b^2+1)}}\, dt \qquad (4.7)$$

The left side of equation (4.7) could be transformed to the proper *analytical* integral [10] in regard to function $\Omega_2$ [15-16] (let us choose the sign "+" in (4.4) for *b*):

$$\int \frac{du}{(u+D)\sqrt{Au^2+Bu+C}} = \frac{\arcsin\left(\frac{(B-2AD)u-BD+2C}{(u+D)\sqrt{B^2-4AC}}\right)}{\sqrt{BD-C-AD^2}}$$

$$\text{if } \{BD > C + AD^2 \text{ and } B^2 > 4AC\} \quad \text{where} \quad A=-b, \quad B=2\Omega_3, \quad C=0, \quad D=2b\Omega_3, \quad \Rightarrow$$

$$\Rightarrow \quad 1)\ 4b(\Omega_3)^2 > -4b^2(\Omega_3)^2\ (\Rightarrow b > -b^2), \quad \text{and} \quad 2)\ 4(\Omega_3)^2 > 0 \qquad \Rightarrow$$

$$\int \frac{d\Omega_2}{(\Omega_2+2b\Omega_3)\sqrt{2(\Omega_2\Omega_3)-(\Omega_2)^2 b}} = \frac{\arcsin\left(\frac{(1+2b^2)\Omega_2-2b\Omega_3}{(\Omega_2+2b\Omega_3)}\right)}{2\Omega_3\sqrt{b+b^3}} \qquad (4.8)$$

**Appendix, A2 (transforming equation (5.5)).**

Let us simplify or transform equation (5.5) to obtain the appropriate expression for the dynamics of angle $\psi$:

$$\frac{d}{dt}\left(\arctan\left(\sqrt{\frac{1-\sin(2\Omega_3 t)}{1+\sin(2\Omega_3 t)}}\right)\right) =$$

$$= \frac{\dfrac{1}{2\sqrt{\dfrac{1-\sin(2\Omega_3 t)}{1+\sin(2\Omega_3 t)}}}}{1+\left(\dfrac{1-\sin(2\Omega_3 t)}{1+\sin(2\Omega_3 t)}\right)}\left(\frac{-2\Omega_3\cos(2\Omega_3 t)(1+\sin(2\Omega_3 t))-2\Omega_3\cos(2\Omega_3 t)(1-\sin(2\Omega_3 t))}{(1+\sin(2\Omega_3 t))^2}\right)$$

$$\Rightarrow \quad \frac{d}{dt}\left(\arctan\left(\sqrt{\frac{1-\sin(2\Omega_3 t)}{1+\sin(2\Omega_3 t)}}\right)\right) = \frac{\dfrac{1}{\sqrt{\dfrac{1-\sin(2\Omega_3 t)}{1+\sin(2\Omega_3 t)}}}}{1+\left(\dfrac{1-\sin(2\Omega_3 t)}{1+\sin(2\Omega_3 t)}\right)}\left(\frac{-2\Omega_3\cos(2\Omega_3 t)}{(1+\sin(2\Omega_3 t))^2}\right)$$

$$\Rightarrow \quad \frac{d}{dt}\left(\arctan\left(\sqrt{\frac{1-\sin(2\Omega_3 t)}{1+\sin(2\Omega_3 t)}}\right)\right) = \frac{\left(\dfrac{1}{\sqrt{(1-\sin(2\Omega_3 t))(1+\sin(2\Omega_3 t))}}\right)}{(1+\sin(2\Omega_3 t))+(1-\sin(2\Omega_3 t))}\left(-2\Omega_3\cos(2\Omega_3 t)\right)$$

$$\Rightarrow \frac{d}{dt}\left(\arctan\left(\sqrt{\frac{1-\sin(2\Omega_3 t)}{1+\sin(2\Omega_3 t)}}\right)\right) = \frac{\left(\dfrac{\left(-2\Omega_3\cos(2\Omega_3 t)\right)}{\sqrt{1-\sin^2(2\Omega_3 t)}}\right)}{2} = -\frac{\left(\dfrac{2\Omega_3\cos(2\Omega_3 t)}{\mp\cos(2\Omega_3 t)}\right)}{2} = \pm\,\Omega_3$$

$$\Rightarrow \qquad \dot{\psi} = \frac{\Omega_3-\dot{\varphi}}{\gamma_3} \quad \Rightarrow \quad \dot{\psi} = \frac{\Omega_3+\dfrac{d}{dt}\left(\arctan\left(\sqrt{\dfrac{1-\sin(2\Omega_3 t)}{1+\sin(2\Omega_3 t)}}\right)\right)}{\sigma\left(\dfrac{1+\sin(2\Omega_3 t)-2b^2}{1+\sin(2\Omega_3 t)+2b^2}\right)},$$

where we choose one of two possible solution (according to the sign "-" in the term $\pm\,\Omega_3$) as below

$$\Rightarrow \quad \dot{\psi} \;=\; 0 \quad \Rightarrow \quad \psi \;=\; const \qquad (5.5)$$

## Conflict of interest

Authors declare that there is no conflict of interests regarding publication of article